\documentclass[aps,prd,onecolumn,groupedaddress,showpacs,nofootinbib,amssymb]{revtex4}
\usepackage[dvips]{graphicx}
\usepackage{amssymb}
\usepackage{amsmath}
\usepackage{hyperref}
\usepackage{graphicx,color}
\usepackage{amsfonts}
\usepackage{bm}
\usepackage{cancel}
\usepackage{comment}

\newcommand\be{\begin{equation}}
\newcommand\ee{\end{equation}}

\allowdisplaybreaks[4]

\begin{document}

\tolerance=5000

\title{Phenomenology of Axionic Static Neutron Stars with Masses in the Mass-Gap Region}
\author{V.K. Oikonomou,$^{1,2}$}
\email{voikonomou@gapps.auth.gr,v.k.oikonomou1979@gmail.com}
\affiliation{$^{1)}$Department of Physics, Aristotle University of
Thessaloniki, Thessaloniki 54124, Greece\\
 $^{2)}$L.N. Gumilyov
Eurasian National University - Astana, 010008, Kazakhstan}

\tolerance=5000

\begin{abstract}
In this work we consider an axionic scalar-tensor theory of
gravity and its effects on static neutron stars. The axionic
theory is considered in the regime in which the axion oscillates
around its potential minimum, which cosmologically occurs
post-inflationary, when the Hubble rate is of the same order as
the axion mass. We construct the TOV equations for this axionic
theory and for a spherically symmetric static spacetime and we
solve these numerically using a quite robust double shooting LSODA
based python integration method. Regarding the equations of state,
we used nine mainstream and quite popular ones, namely, the WFF1,
the SLy, the APR, the MS1, the AP3, the AP4, the ENG, the MPA1 and
the MS1b, using the piecewise polytropic description for each.
From the extracted data we calculate the Jordan frame masses and
radii, and we confront the resulting phenomenology with five
well-known neutron star constraints. As we demonstrate, the AP3,
the ENG and the MPA1 equations of state yield phenomenologically
viable results which are compatible with the constraints, with the
MPA1 equation of state enjoying an elevated role among the three.
The reason is that the MPA1 fits well the phenomenological
constraints. A mentionable feature is the fact that all the viable
phenomenologically equations of state produce maximum masses which
are in the mass-gap region with $M_{max}>2.5M_{\odot}$, but lower
that the causal 3 solar masses limit. We also compare the neutron
star phenomenology produced by the axionic scalar-tensor theory
with the phenomenology produced by inflationary attractors
scalar-tensor theories.
\end{abstract}


\maketitle

\section*{Introduction}

\begin{figure}[h!]
\centering
\includegraphics[width=30pc]{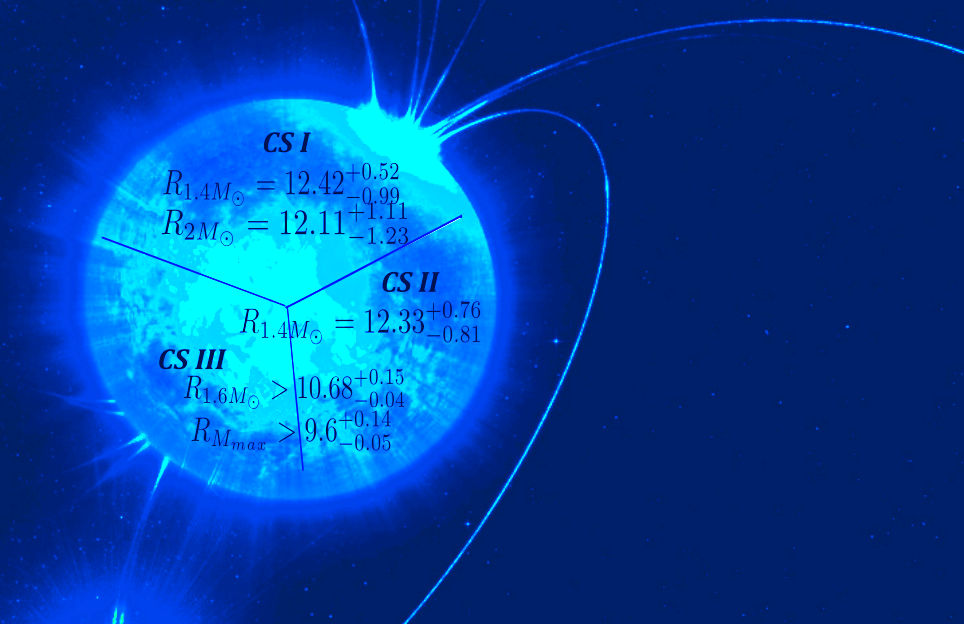}
\caption{Pictorial presentation of the constraints CSI
\cite{Altiparmak:2022bke} $R_{1.4M_{\odot}}=12.42^{+0.52}_{-0.99}$
and $R_{2M_{\odot}}=12.11^{+1.11}_{-1.23}\,$km, CSII
\cite{Raaijmakers:2021uju} with
$R_{1.4M_{\odot}}=12.33^{+0.76}_{-0.81}\,\mathrm{km}$ and CSIII
\cite{Bauswein:2017vtn} according to which the radius of an
$1.6M_{\odot}$ mass NS must be
$R_{1.6M_{\odot}}>10.68^{+0.15}_{-0.04}\,$km while for NSs with
the maximum mass, the radius must be
$R_{M_{max}}>9.6^{+0.14}_{-0.03}\,$km. This figure is edited and
it is based on a public image of ESO, which can be found freely in
Credit: ESO/L.Cal\c{c}ada:
\url{https://www.eso.org/public/images/eso0831a/.}} \label{plotcs}
\end{figure}

The current focus of theoretical astrophysicists, cosmologists,
the theoretical particle physicists and astronomers lies in the
sky, where current and future observations are anticipated and are
expected to verify current theories or shake the ground in
theoretical astrophysics and cosmology. The striking chorus of the
new physics era observations initiated with the kilonova GW170817
event observation back in 2017
\cite{TheLIGOScientific:2017qsa,Abbott:2020khf}. This was a
remarkable event exactly because a kilonova was involved in the
observation. Thus astronomers of the LIGO-Virgo collaboration
detected a gravitational wave and an electromagnetic wave emitted
by the two neutron star (NS) merger. The two waves arrived almost
simultaneously, since the electromagnetic wave arrived 1.74
seconds after the gravitational wave, and thus this event, almost
instantly after the GW170817 announcement, casted serious doubts
on the viability of theories of gravity that predict a
gravitational wave speed different from that of light's, see for
example
\cite{Ezquiaga:2017ekz,Baker:2017hug,Creminelli:2017sry,Sakstein:2017xjx},
although theoretical refinements have been proposed for theories
with stringy origin
\cite{Odintsov:2020sqy,Oikonomou:2021kql,Oikonomou:2022ksx}. Many
exciting events have been observed after the GW170817 event, with
milestones being the GW190814 event \cite{LIGOScientific:2020zkf}
and the striking 2023 NANOGrav observation of a stochastic
gravitational wave background, also confirmed by other Pulsar
Timing Array (PTA) collaborations
\cite{nanograv,Antoniadis:2023ott,Reardon:2023gzh,Xu:2023wog}. The
former event captured the merging event of a black hole with a
mysterious small mass object in the mass gap region $M\sim
2.6\,M_{\odot}$, which could be the heaviest NS or the lightest
black hole, but in both cases the outcome is exciting to say the
least. To date nobody confirmed either of the aforementioned
scenarios. Apparently the GW190814 event was a very valuable
present offered by Nature to us, and can be the first event that
gives us a hint for the existence of heavy NSs which cannot be
explained by General Relativity (GR) alone, even with the stiffest
equation of state (EoS). What is needed to confirm the existence
of such heavy NSs is either a NS merger event accompanied by a
kilonova, or some observation of an low-spin isolated pulsar like
the black-widow binary pulsar PSR J0952-0607 which has mass
$M=2.35\pm 0.17$ \cite{Romani:2022jhd} and it is quite close to
the mass gap region. Motivated by this line of reasoning it stands
to reason to think that GR itself might not suffice to describe
heavy NSs in the mass gap region and that some modified gravity
\cite{reviews1,reviews2,reviews3,reviews4,reviews5} can actually
describe heavy NSs. This is a long-standing question in
theoretical astrophysics and it is our personal belief that this
question will be answered by a ground breaking near future
kilonova NSs merger event. Thus NSs
\cite{Haensel:2007yy,Friedman:2013xza,Baym:2017whm,Lattimer:2004pg,Olmo:2019flu}
are expected to be the test-bed of theories in the near future and
for the next generations of scientists. There is a lot of physics
involved in NSs and many theories can be experimentally tested,
like for example nuclear physics theories with extreme matter
conditions
\cite{Lattimer:2012nd,Steiner:2011ft,Horowitz:2005zb,Watanabe:2000rj,Shen:1998gq,Xu:2009vi,Hebeler:2013nza,Mendoza-Temis:2014mja,Ho:2014pta,Kanakis-Pegios:2020kzp,Tsaloukidis:2022rus,Kanakis-Pegios:2023gvc},
high energy theoretical and particle physics theories
\cite{Buschmann:2019pfp,Safdi:2018oeu,Hook:2018iia,Edwards:2020afl,Nurmi:2021xds},
modified gravity descriptions of NSs,
\cite{Astashenok:2020qds,Astashenok:2021peo,Capozziello:2015yza,Astashenok:2014nua,Astashenok:2014pua,Astashenok:2013vza,Arapoglu:2010rz,Panotopoulos:2021sbf,Lobato:2020fxt,Numajiri:2021nsc},
see also
\cite{Pani:2014jra,Staykov:2014mwa,Horbatsch:2015bua,Silva:2014fca,Doneva:2013qva,Xu:2020vbs,Salgado:1998sg,Shibata:2013pra,Arapoglu:2019mun,Ramazanoglu:2016kul,AltahaMotahar:2019ekm,Chew:2019lsa,Blazquez-Salcedo:2020ibb,Motahar:2017blm,Odintsov:2021qbq,Odintsov:2021nqa,Oikonomou:2021iid,Pretel:2022rwx,Pretel:2022plg,Cuzinatto:2016ehv,Oikonomou:2023dgu,Odintsov:2023ypt,Oikonomou:2023lnh}
for the scalar-tensor approach and of course theoretical
astrophysics scenarios,
\cite{Altiparmak:2022bke,Bauswein:2020kor,Vretinaris:2019spn,Bauswein:2020aag,Bauswein:2017vtn,Most:2018hfd,Rezzolla:2017aly,Nathanail:2021tay,Koppel:2019pys,Raaijmakers:2021uju,Most:2020exl,Ecker:2022dlg,Jiang:2022tps,Biswas:2023ceq,Liodis:2023adg}.
Motivated by the importance of the modified gravity perspective in
NSs, in this work we shall examine static NSs in the context of an
axionic scalar field theory. The axion is of profound theoretical
importance and it is considered as an important dark matter
candidate nowadays, and for a mainstream of important articles and
reviews on axions see
\cite{Preskill:1982cy,Abbott:1982af,Dine:1982ah,Marsh:2015xka,Sikivie:2006ni,Co:2019jts,Co:2020dya,Chen:2022nbb,Oikonomou:2022ela,Roy:2021uye,Tsai:2021irw,Visinelli:2018utg,Oikonomou:2022tux,Odintsov:2020iui,Oikonomou:2020qah,Vagnozzi:2022moj,Banerjee:2021oeu,Machado:2019xuc,Machado:2018nqk,Heinze:2023nfb}.
We shall consider several EoSs, using a phenomenological piecewise
polytropic approach \cite{Read:2008iy,Read:2009yp}, which is more
appropriate for phenomenological reasons. With regard to the EoSs
we shall consider the following, the SLy \cite{Douchin:2001sv},
the AP3-AP4 \cite{Akmal:1998cf}, the WFF1 \cite{Wiringa:1988tp},
the ENG \cite{Engvik:1995gn}, the MPA1 \cite{Muther:1987xaa}, the
MS1 and MS1b \cite{Mueller:1996pm} and finally the APR EoS
\cite{Akmal:1997ft}. Among all these EoSs, the MPA1 seems to
produce extremely viable NS phenomenology as it was shown in Ref.
\cite{Odintsov:2023ypt}. We tried the most important
phenomenologically EoSs, which include stiff and mildly stiff
EoSs, and the reason is that we wanted to check the variety of
phenomenological implications. After our examination we found that
some EoSs provide extreme maximum masses but these EoSs are
excluded phenomenologically. Remarkably all the EoSs that provide
viable results yield maximum NSs masses that respect the causal
maximum mass of general relativity. Thus the inclusion of so many
EoSs was from curiosity and for phenomenological completeness. We
shall numerically solve the Einstein frame
Tolman-Oppenheimer-Volkoff (TOV) equations and we shall calculate
the Jordan frame Arnowitt-Deser-Misner (ADM) gravitational mass
and radius of the NS \cite{Arnowitt:1960zzc}. The phenomenological
viability of the NS will be tested by confronting the
gravitational mass and radius data with all the current
constraints on the masses and radii of NS. Specifically we shall
consider three types of constraints which we call CSI, CSII and
CSIII along with the latest NICER constraints. Specifically, the
constraint CSI \cite{Altiparmak:2022bke} constrains the radius of
a $1.4M_{\odot}$ mass NS
 to be
$R_{1.4M_{\odot}}=12.42^{+0.52}_{-0.99}$ while that of an
$2M_{\odot}$ mass NS must be
$R_{2M_{\odot}}=12.11^{+1.11}_{-1.23}\,$km. The constraint CSII
\cite{Raaijmakers:2021uju} and constrains the radius of a
$1.4M_{\odot}$ mass NS to be
$R_{1.4M_{\odot}}=12.33^{+0.76}_{-0.81}\,\mathrm{km}$, while CSIII
\cite{Bauswein:2017vtn} constrains the radius of an $1.6M_{\odot}$
mass NS to be larger than
$R_{1.6M_{\odot}}>10.68^{+0.15}_{-0.04}\,$km, and also the radius
corresponding to the maximum mass of the NS must be larger than
$R_{M_{max}}>9.6^{+0.14}_{-0.03}\,$km. We have gathered the
constraints CSI, CSII and CSIII in Fig. \ref{plotcs}. Regarding
the NICER constraints, we shall consider two, which constrain the
radius of an $M=1.4M_{\odot}$ mass NS to be
$R_{1.4M_{\odot}}=11.34-13.23\,$km \cite{Miller:2021qha}, which we
call NICER I, while the second NICER constraint, to which we shall
refer as NICER II, constraints again the radius of a
$M=1.4M_{\odot}$ mass NS to be $R_{1.4M_{\odot}}=12.33-13.25\,$km.
These are included in Table \ref{table0}.
\begin{table}[h!]
  \begin{center}
    \caption{\emph{\textbf{Viability Constraints for NS Phenomenology}}}
    \label{table0}
    \begin{tabular}{|r|r|}
     \hline
      \textbf{Constraint}   & \textbf{Mass and Radius}
      \\ \hline
      \textbf{CSI} & For $M=1.4M_{\odot}$, $R_{1.4M_{\odot}}=12.42^{+0.52}_{-0.99}$ and for $M=2M_{\odot}$, $R_{2M_{\odot}}=12.11^{+1.11}_{-1.23}\,$km.
\\  \hline
      \textbf{CSII} & For $M=1.4M_{\odot}$, $R_{1.4M_{\odot}}=12.33^{+0.76}_{-0.81}\,\mathrm{km}$.
\\  \hline
\textbf{CSIII} & For $M=1.6M_{\odot}$,
$R_{1.6M_{\odot}}>10.68^{+0.15}_{-0.04}\,$km, and for $M=M_{max}$,
$R_{M_{max}}>9.6^{+0.14}_{-0.03}\,$km.
\\  \hline
 NICER I & For $M=1.4M_{\odot}$, $11.34\,\mathrm{km}<R_{1.4M_{\odot}}<13.23\,\mathrm{km}$
\\  \hline
 NICER II & For $M=1.4M_{\odot}$,
 $12.33\,\mathrm{km}<R_{1.4M_{\odot}}<13.25\,\mathrm{km}$
\\  \hline
    \end{tabular}
  \end{center}
\end{table}
After presenting the phenomenological outcomes of our study we
shall answer the theoretical question whether NS phenomenology
originating by scalar-tensor inflationary potentials and axionic
potentials NS produce similar results. The answer, to our
surprise, lies in the affirmative. The surprise is due to the fact
that although we use a natural inflation potential (axionic) the
approximations used for this axion potential correspond to an era
where the axion oscillates around the minimum of its potential and
cosmologically redshifts as dark matter. We expected some
differences, but the result is that the phenomenology of NSs is
similar in the two cases. Now it is important to note that the
axion potential in cosmological scales is basically a theory of
gravity active at large scales, so the question is what is the
relevance of this cosmological theory. The answer to this is that
NSs are basically extreme gravitational environments, so in
principle the large scale gravitational theory might have direct
effects on such extreme gravity environments. Also we need to
stress that the axionic NSs is composed by ordinary matter obeying
one of the EoSs mentioned previously, but the gravitational
equilibrium is controlled by the axionic scalar-tensor theory,
thus it affects the maximum mass and the radius of the NS. Also
note that we did not assume that actual dark matter particles,
such as axions, exist inside the core of the NSs. This is a quite
interesting perspective, but the exact treatment which somewhat
change the theoretical framework used, since it would probably
change the hydrodynamic equilibrium of the NS to some extent.

\section{Neutron Stars Physics and Scalar-tensor Field Theories}

Let us review in brief the formalism of scalar-tensor theories in
the Einstein frame, and also extract the gravitational mass of NSs
in the Einstein frame. We use the notation of Ref.
\cite{Pani:2014jra} and we also shall work in Geometrized units
($G=c=1$). The Jordan frame action of a non-minimally coupled
scalar field is,
\begin{equation}\label{taintro}
\mathcal{S}=\int
d^4x\frac{\sqrt{-g}}{16\pi}\Big{[}\Omega(\phi)R-\frac{1}{2}g^{\mu
\nu}\partial_{\mu}\phi\partial_{\nu}\phi-U(\phi)\Big{]}+S_m(\psi_m,g_{\mu
\nu})\, ,
\end{equation}
and upon conformally transforming this action, transformation,
\begin{equation}\label{ta1higgsintro}
\tilde{g}_{\mu \nu}=A^{-2}g_{\mu \nu}\,
,\,\,\,A(\phi)=\Omega^{-1/2}(\phi)\, ,
\end{equation}
we get the Einstein frame action,
\begin{equation}\label{ta5higgsintro}
\mathcal{S}=\int
d^4x\sqrt{-\tilde{g}}\Big{(}\frac{\tilde{R}}{16\pi}-\frac{1}{2}
\tilde{g}_{\mu \nu}\partial^{\mu}\varphi
\partial^{\nu}\varphi-\frac{V(\varphi)}{16\pi}\Big{)}+S_m(\psi_m,A^2(\varphi)\tilde{g}_{\mu
\nu})\, ,
\end{equation}
with $\varphi$ denoting the Einstein frame, which has a scalar
potential $V(\varphi)$ related to the Jordan frame one $U(\phi)$
in the following way,
\begin{equation}\label{potentialns1intro}
V(\varphi)=\frac{U(\phi)}{\Omega^2}\, .
\end{equation}
Note that the passage from the Jordan to the Einstein frame is
always possible in scalar-tensor theories, since a non-diverging
conformal transformation can always be found. The passage from the
Einstein-frame to the Jordan frame is not possible in higher
derivative gravities that contain terms of the Gauss-Bonnet
invariant, the Riemann and Ricci tensors. Furthermore, an
important function is $\alpha(\varphi)$ which is defined as
follows,
\begin{equation}\label{alphaofvarphigeneraldefintro}
\alpha(\varphi)=\frac{d \ln A(\varphi)}{d \varphi}\, ,
\end{equation}
which is essential for the TOV equations and also
$A(\varphi)=\Omega^{-1/2}(\phi)$.
Since we consider static NSs, we shall consider a spherically
symmetric metric,
\begin{equation}\label{tov1intro}
ds^2=-e^{\nu(r)}dt^2+\frac{dr^2}{1-\frac{2
m(r)}{r}}+r^2(d\theta^2+\sin^2\theta d\phi^2)\, ,
\end{equation}
with $m(r)$ describing the gravitational mass of the NS, and $r$
is the circumferential radius. The aim of the numerical solution
we shall provide is the determination of the functions $\nu(r)$
and $\frac{1}{1-\frac{2 m(r)}{r}}$, which in the case of
scalar-tensor gravity also receive contributions beyond the
surface of the NSs, a feature absent in the GR case, and this is
due to the fact that the scalar field affects these two metric
functions beyond the surface of the star. Hence, there is no
conventional matching of the spherically symmetric metric with the
Schwarzschild metric at the surface of the star. This matching is
performed at numerical infinity, where the effects of the scalar
field have been smoothed away. Now by varying the Einstein frame
action in the presence of ordinary matter, we get the TOV
equations,
\begin{equation}\label{tov2intro}
\frac{d m}{dr}=4\pi r^2
A^4(\varphi)\varepsilon+\frac{r}{2}(r-2m(r))\omega^2+4\pi
r^2V(\varphi)\, ,
\end{equation}
\begin{equation}\label{tov3intro}
\frac{d\nu}{dr}=r\omega^2+\frac{2}{r(r-2m(r))}\Big{[}4\pi
A^4(\varphi)r^3P-4\pi V(\varphi)
r^3\Big{]}+\frac{2m(r)}{r(r-2m(r))}\, ,
\end{equation}
\begin{equation}\label{tov4intro}
\frac{d\omega}{dr}=\frac{4\pi r
A^4(\varphi)}{r-2m(r)}\Big{(}\alpha(\varphi)(\epsilon-3P)+
r\omega(\epsilon-P)\Big{)}-\frac{2\omega
(r-m(r))}{r(r-2m(r))}+\frac{8\pi \omega r^2 V(\varphi)+r\frac{d
V(\varphi)}{d \varphi}}{r-2 m(r)}\, ,
\end{equation}
\begin{equation}\label{tov5intro}
\frac{dP}{dr}=-(\epsilon+P)\Big{[}\frac{1}{2}\frac{d\nu}{dr}+\alpha
(\varphi)\omega\Big{]}\, ,
\end{equation}
\begin{equation}\label{tov5newfinalintro}
\omega=\frac{d \varphi}{dr}\, ,
\end{equation}
with $\alpha (\varphi)$ being defined in Eq.
(\ref{alphaofvarphigeneraldefintro}). For the numerical analysis
we use the following initial conditions,
\begin{equation}\label{tov8intro}
P(0)=P_c\, ,\,\,\,m(0)=0\, , \,\,\,\nu(0)=-\nu_c\, ,
\,\,\,\varphi(0)=\varphi_c\, ,\,\,\, \omega (0)=0\, ,
\end{equation}
with the initial values $\nu_c$ and $\varphi_c$ being initially
arbitrary, but the exact correct value will be obtained by
performing a double shooting method, which aims to find the
optimal values which render the scalar field at numerical infinity
to be zero. Regarding the EoS for the nuclear matter, we shall use
nine piecewise polytropic \cite{Read:2008iy,Read:2009yp} EoSs, and
specifically, the SLy \cite{Douchin:2001sv}, the AP3-AP4
\cite{Akmal:1998cf}, the WFF1 \cite{Wiringa:1988tp}, the ENG
\cite{Engvik:1995gn}, the MPA1 \cite{Muther:1987xaa}, the MS1 and
MS1b \cite{Mueller:1996pm} and finally the APR EoS
\cite{Akmal:1997ft}. Let us now present the formula for the
Einstein frame ADM mass of the NS, so we introduce the quantities
$K_E$ and $K_J$ defined as follows,
\begin{equation}\label{hE}
\mathcal{K}_E=1-\frac{2 m}{r_E}\, ,
\end{equation}
\begin{equation}\label{hE}
\mathcal{K}_J=1-\frac{2  m_J}{r_J}\, ,
\end{equation}
and are related as follows,
\begin{equation}\label{hehjrelation}
\mathcal{K}_J=A^{-2}\mathcal{K}_E\, .
\end{equation}
Accordingly, the Jordan and Einstein frame radii are related as
follows,
\begin{equation}\label{radiiconftrans}
r_J=A r_E\, ,
\end{equation}
and accordingly the Jordan frame ADM gravitational mass is,
\begin{equation}\label{jordaframemass1}
M_J=\lim_{r\to \infty}\frac{r_J}{2}\left(1-\mathcal{K}_J \right)
\, ,
\end{equation}
while the Einstein frame ADM gravitational mass is,
\begin{equation}\label{einsteiframemass1}
M_E=\lim_{r\to \infty}\frac{r_E}{2}\left(1-\mathcal{K}_E \right)
\, .
\end{equation}
Asymptotically Eq. (\ref{hehjrelation}) yields,
\begin{equation}\label{asymptotich}
\mathcal{K}_J(r_E)=\left(1+\alpha(\varphi(r_E))\frac{d \varphi}{d
r}r_E \right)^2\mathcal{K}_E(\varphi(r_E))\, ,
\end{equation}
with $r_E$ denoting the Einstein frame radius at numerical
infinity and also $\frac{d\varphi }{dr}=\frac{d\varphi
}{dr}\Big{|}_{r=r_E}$. Combining Eqs.
(\ref{hE})-(\ref{asymptotich}) we get the formula for the Jordan
frame ADM gravitational mass of the NS,
\begin{equation}\label{jordanframeADMmassfinal}
M_J=A(\varphi(r_E))\left(M_E-\frac{r_E^{2}}{2}\alpha
(\varphi(r_E))\frac{d\varphi
}{dr}\left(2+\alpha(\varphi(r_E))r_E\frac{d \varphi}{dr}
\right)\left(1-\frac{2 M_E}{r_E} \right) \right)\, ,
\end{equation}
and recall that $\frac{d\varphi }{dr}=\frac{d\varphi
}{dr}\Big{|}_{r=r_E}$.
\begin{figure}[h!]
\centering
\includegraphics[width=18pc]{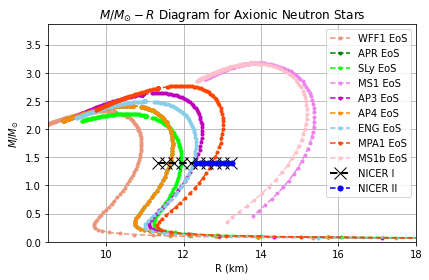}
\caption{The $M-R$ graphs for the axionic NSs for the EoSs WFF1,
SLy, APR, MS1, AP3, AP4, ENG, MPA1, MS1b confronted with
constraints NICER I \cite{Miller:2021qha} and NICER II
\cite{Ecker:2022dlg}.} \label{plot1}
\end{figure}
Also the radius of the NS in the Jordan frame denoted as $R$ is
related to the Einstein frame one $R_s$ as follows,
\begin{equation}\label{radiussurface}
R=A(\varphi(R_s))\, R_s\, ,
\end{equation}
and note that the Einstein frame mass is determined at the surface
of the star where $P(R_s)=0$. Our aim with the numerical analysis
is to calculate the Einstein frame mass and radii of the NS and
from these to calculate their Jordan frame counterparts and
construct the $M-R$ graphs of the NSs.

\subsection{Axion Neutron Stars}

Now let us consider the axionic scalar-tensor theory which we
shall assume that governs the Universe at cosmological and
astrophysical scales. The axionic theory is basically a dark
matter theory. Let us consider the general form of such theory in
a cosmological context, and then we specify the theory for the NS
study we aim to perform in this work. In the context of the
misalignment axion \cite{Marsh:2015xka,Co:2019jts}, the axion
possesses a primordial $U(1)$ Peccei-Quinn symmetry which is
basically broken during the inflationary era. The misalignment
axion commences its evolution towards the minimum of its scalar
potential which has the following form,
\begin{equation}\label{axionpotentnialfull}
V_a(\phi )=m_a^2f_a^2\left(1-\cos \left(\frac{\phi}{f_a}\right)
\right)\, ,
\end{equation}
where initially its initial value $\phi_i$ at the time it
commences its motion towards the bottom of its potential is
$\phi_i\sim f_a$, with $f_a$ being the axion decay constant, which
is of the order $f_a>10^{9}\,$GeV. During its motion towards the
minimum of the potential the axion satisfies, $\phi/f_a<1$, thus
for the original axionic theory, the scalar potential can be
approximated by,
\begin{equation}\label{axionpotential}
V_a(\phi )\simeq \frac{1}{2}m_a^2\phi^2\, ,
\end{equation}
an approximation which is valid when $\phi< f_a$ and this covers
the eras in which the axion moves towards the minimum of the
scalar potential. After the axion reaches the minimum it commences
coherent oscillations, approximately when the Hubble rate of the
Universe $H$ is of the order as the axion mass $m_a$, so basically
when $H\sim m_a$. Note that this occurred primordially, since the
present day Hubble rate is $H\sim 10^{-33}\,$eV so it is too small
to be compared to the axion mass, which in most cases, the axion
mass is in he range $m_a\sim 10^{-10}-10^{-24}\,$eV. After this
era primordial era with $H\sim m_a$, the axion oscillates and its
energy density redshifts as cold dark matter and the axion
basically becomes a non-thermal dark matter candidate.
\begin{figure}[h!]
\centering
\includegraphics[width=18pc]{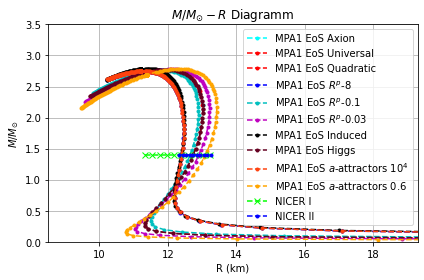}
\caption{The $M-R$ graphs for the axionic NSs and of most
well-known inflationary attractors, for the MPA1 EoS versus the
NICER I \cite{Miller:2021qha} and NICER II \cite{Ecker:2022dlg}
constraints.} \label{plot2}
\end{figure}
For the purposes of this work, we shall consider the non-minimally
coupled Jordan frame axionic theory developed in Ref.
\cite{Reyimuaji:2020goi}, in which the Jordan frame potential of
Eq. (\ref{taintro}) in Geometrized units reads,
\begin{equation}\label{alphaattractpotensalphaattractors}
U(\phi)=\Lambda^4\left(1-\cos\left(\frac{\phi}{f_a} \right)
\right)\, ,
\end{equation}
while the non-minimal coupling function $\Omega(\phi)$ in Eq.
(\ref{taintro}) reads,
\begin{equation}\label{nonminimalcoupling}
\Omega(\phi)=1+\xi \phi^n\, .
\end{equation}
The parameter $n$ can take values $n=2,4,...$ so we focus here to
$n=2$. The parameter $\xi$ shall be assumed to take large values
$\xi\sim 10^4\gg 1$. Then we get,
\begin{equation}\label{finalvarphiphialphaattractors}
\frac{d \varphi }{d
\phi}=\sqrt{\frac{1}{4\pi}}\sqrt{\frac{12\xi^2\phi^2+1\xi\phi^2}{4\left(1+\xi\phi^2\right)^2}}\,
,
\end{equation}
so assuming that in the interior of the star, the following
approximation holds true,
\begin{equation}\label{approximationaxion}
\xi^2\phi^2\gg \xi \phi^2\gg 1\, ,
\end{equation}
Eq. (\ref{finalvarphiphialphaattractors}) becomes,
\begin{equation}\label{finalvarphiphialphaattractorsaxion}
\frac{d \varphi }{d \phi}=\sqrt{\frac{3}{4\pi}}\frac{\xi
\phi}{1+\xi\phi^2}\, ,
\end{equation}
so upon integration of the above we get,
\begin{equation}\label{fvarphialphaattractors}
\varphi=\frac{\sqrt{3}}{2\sqrt{4\pi}}\ln \left(1+\xi
\phi^2\right)\, ,
\end{equation}
or equivalently,
\begin{equation}\label{equivrelation}
1+\xi \phi^2=e^{2\sqrt{\frac{4\pi}{3}}\varphi}\, .
\end{equation}
The conformal factor $A(\phi)$ expressed in terms of the Einstein
frame scalar field $\varphi$ reads,
\begin{equation}\label{Aofpvarphiprofinalalphaattractors}
A(\varphi)=e^{-\sqrt{\frac{4\pi}{3}}\varphi}\, ,
\end{equation}
and the function  $\alpha(\varphi)$ which is defined in Eq.
(\ref{alphaofvarphigeneraldefintro}) takes the following form,
\begin{equation}\label{alphaofphifinalintermsofvarphialphaattractors}
a(\varphi)=\alpha=-\sqrt{\frac{4\pi}{3}}\, .
\end{equation}
\begin{table}[h!]
  \begin{center}
    \caption{\emph{\textbf{Maximum Masses for Axionic NSs in the Mass Gap Region.}}}
    \label{tablemaxmasses}
    \begin{tabular}{|r|r|r|r|r|}
     \hline
      \textbf{Model}   & \textbf{MPA1 EoS} & \textbf{MS1b EoS} &
      \textbf{AP3 EoS} & \textbf{MS1 EoS}
      \\  \hline
      \textbf{$M_{MAX}$} & $M_{MPA1}= 2.771\,M_{\odot}$ & $M_{MS1b}= 3.167\, M_{\odot}$ & $M_{AP3}= 2.638\,
M_{\odot}$ & $M_{MS1}= 3.175\,M_{\odot}$
\\  \hline
    \end{tabular}
  \end{center}
\end{table}
Finally, since we are interested in an era in which $\phi/f_a\ll
1$, the Einstein frame potential takes the following form,
\begin{equation}\label{einsteinframepotentialfinalalphaattractors}
V(\varphi)=-\frac{\Lambda^4}{2f_a\xi}\left(1-e^{-\alpha \varphi}
\right)e^{-2\alpha\varphi}\, ,
\end{equation}
where we expanded the cosine term in the Jordan frame in the limit
$\phi/f_a\ll 1$. Now regarding the free parameters, the axion mass
determines the axion mass so in natural units it must be
$\Lambda^4/f_a^2\sim m_a^2$ so in order to have an axion with mass
$m_a\sim 10^{-10}\,$eV with $f_a\sim 10^9\,$GeV we have
$\Lambda\sim 10^4\,$eV. Also
$\xi>-\left(\frac{M_p}{f_a\pi}\right)^2$ where $M_p$ is the
reduced Planck mass in natural units. In Geometrized units the
parameter $\xi$ must be $\xi\sim 10^4$. In the end of the
calculations, the approximation (\ref{approximationaxion}) must be
checked if it holds true in the center and at the surface of the
star.
\begin{table}[h!]
  \begin{center}
    \caption{\emph{\textbf{Axionic NSs vs CSI for NS Masses $M\sim 2M_{\odot}$, $R_{2M_{\odot}}=12.11^{+1.11}_{-1.23}\,$km, for the SLy, APR, WFF1, MS1 and AP3 EoSs. The ''x'' denotes non-viability.}}}
    \label{tablecsi2}
    \begin{tabular}{|r|r|r|r|r|r|}
     \hline
      \textbf{Model}   & \textbf{SLy EoS} & \textbf{APR EoS} & \textbf{WFF1
      EoS} & \textbf{MS1 EoS} & \textbf{AP3 EoS}
      \\  \hline
      \textbf{Axionic NSs Radii} & $R_{SLy}= 11.675\,$Km & $R_{APR}=  11.650\,$Km & $R_{WFF1}=
      x$
       & $R_{MS1}= x$ & $R_{AP3}= 12.467\,$Km
\\  \hline
    \end{tabular}
  \end{center}
\end{table}

\begin{table}[h!]
  \begin{center}
    \caption{\emph{\textbf{Axionic NSs vs CSI for NS Masses $M\sim
2M_{\odot}$, $R_{2M_{\odot}}=12.11^{+1.11}_{-1.23}\,$km, for the
AP4, ENG, MPA1 and MS1b. The ''x'' denotes non-viability.}}}
    \label{tablecsi22}
    \begin{tabular}{|r|r|r|r|r|}
     \hline
      \textbf{Model}   & \textbf{AP4 EoS} & \textbf{ENG EoS} &
      \textbf{MPA1 EoS} & \textbf{MS1b EoS}
      \\  \hline
      \textbf{Axionic NSs} & $R_{AP4}= 11.650\,$Km & $R_{ENG}=  12.263\,$Km & $R_{MPA1}=
      13.014\,$Km
       & $R_{MS1b}= x$
\\  \hline
    \end{tabular}
  \end{center}
\end{table}

\subsection{Results and Confrontation with the Data}

In this section we shall present the outcomes of our numerical
analysis. We used a rigorous double shooting LSODA python-based
integration method which is variant of the pyTOV-STT code
developed in Ref. \cite{niksterg}. The double shooting technique
is aimed for finding the optimal initial conditions for $\nu_c$
and $\varphi_c$ defined in Eq. (\ref{tov8intro}) that make the
values of the scalar field at numerical infinity near zero.
\begin{table}[h!]
  \begin{center}
    \caption{\emph{\textbf{Axionic NSs vs CSI for NS Masses $M\sim 1.4M_{\odot}$, $R_{1.4M_{\odot}}=12.42^{+0.52}_{-0.99}$, for the SLy, APR, WFF1, MS1 and AP3 EoSs. The ''x'' denotes non-viability. }}}
    \label{tablecsi14}
    \begin{tabular}{|r|r|r|r|r|r|}
     \hline
      \textbf{Model}   & \textbf{SLy EoS} & \textbf{APR EoS} & \textbf{WFF1
      EoS} & \textbf{MS1 EoS} & \textbf{AP3 EoS}
      \\  \hline
      \textbf{Axionic NSs Radii} & $R_{SLy}= 11.934\,$Km & $R_{APR}=  11.645\,$Km & $R_{WFF1}=
      x$
       & $R_{MS1}= x$ & $R_{AP3}= 12.333$
\\  \hline
    \end{tabular}
  \end{center}
\end{table}
With our numerical analysis we extracted the Einstein frame radii
and gravitational masses for the NSs, using all the nine different
EoSs, and we found the Jordan frame quantities using the formulas
of the previous sections. Using the Jordan frame gravitational
masses and radii we constructed the $M-R$ graphs for all the EoSs
we mentioned earlier. We also took account the constraints CSI,
CSII and CSIII, which are pictorially represented  in Fig.
\ref{plotcs}. Recall that constraint CSI \cite{Altiparmak:2022bke}
indicates that a NS with mass $1.4M_{\odot}$ must have radius
$R_{1.4M_{\odot}}=12.42^{+0.52}_{-0.99}$, while for the case of a
$2M_{\odot}$ mass NS the radius must be
$R_{2M_{\odot}}=12.11^{+1.11}_{-1.23}\,$km. Considering the
constraint CSII \cite{Raaijmakers:2021uju}, for an $1.4M_{\odot}$
mass NS, the radius must be
$R_{1.4M_{\odot}}=12.33^{+0.76}_{-0.81}\,\mathrm{km}$.
\begin{table}[h!]
  \begin{center}
    \caption{\emph{\textbf{Axionic NSs vs CSI for NS Masses $M\sim
1.4M_{\odot}$, $R_{1.4M_{\odot}}=12.42^{+0.52}_{-0.99}$, for the
AP4, ENG, MPA1 and MS1b. The ''x'' denotes non-viability.}}}
    \label{tablecsi142}
    \begin{tabular}{|r|r|r|r|r|}
     \hline
      \textbf{Model}   & \textbf{AP4 EoS} & \textbf{ENG EoS} &
      \textbf{MPA1
      EoS} & \textbf{MS1b EoS}
      \\  \hline
      \textbf{Axionic NSs Radii} & $R_{AP4}= 11.645\,$Km & $R_{ENG}=  12.236\,$Km & $R_{MPA1}=
      12.74\,$Km
       & $R_{MS1b}= x$
\\  \hline
    \end{tabular}
  \end{center}
\end{table}
Finally, considering the CSIII constraint, regarding NSs with
masses $1.6M_{\odot}$, the radius must be
$R_{1.6M_{\odot}}>10.68^{+0.15}_{-0.04}\,$km and also for the
maximum mass of a NS, the radius must be
$R_{M_{max}}>9.6^{+0.14}_{-0.03}\,$km.
\begin{table}[h!]
  \begin{center}
    \caption{\emph{\textbf{Axionic NSs Radii vs CSII for NS Masses $M\sim 1.4M_{\odot}$, $R_{1.4M_{\odot}}=12.33^{+0.76}_{-0.81}\,\mathrm{km}$, for the SLy, APR, WFF1, MS1 and AP3 EoSs. The ''x'' denotes non-viability.}}}
    \label{tablecsii14}
    \begin{tabular}{|r|r|r|r|r|r|}
     \hline
      \textbf{Model}   & \textbf{SLy EoS} & \textbf{APR EoS} & \textbf{WFF1
      EoS} & \textbf{MS1 EoS} & \textbf{AP3 EoS}
      \\  \hline
      \textbf{Axionic NSs Radii} & $R_{SLy}= 11.934\,$Km & $R_{APR}=  11.645\,$Km & $R_{WFF1}=
      x$
       & $R_{MS1}= x$ & $R_{AP3}= 12.287\,$Km
\\  \hline
    \end{tabular}
  \end{center}
\end{table}
\begin{table}[h!]
  \begin{center}
    \caption{\emph{\textbf{Axionic NSs vs CSII for NS Masses $M\sim
1.4M_{\odot}$,
$R_{1.4M_{\odot}}=12.33^{+0.76}_{-0.81}\,\mathrm{km}$, for the
AP4, ENG, MPA1 and MS1b. The ''x'' denotes non-viability.}}}
    \label{tablecsii142}
    \begin{tabular}{|r|r|r|r|r|}
     \hline
      \textbf{Model}   & \textbf{AP4 EoS} & \textbf{ENG EoS} &
      \textbf{MPA1 EoS} & \textbf{MS1b EoS}\\  \hline
      \textbf{Axionic NSs Radii} & $R_{AP4}= 11.645\,$Km & $R_{ENG}=  12.236\,$Km & $R_{MPA1}=
      12.748\,$Km  & $R_{MS1b}= x$
\\  \hline
    \end{tabular}
  \end{center}
\end{table}
In all the $M-R$ graphs we considered, we also included the NICER
I constraint regarding $M=1.4M_{\odot}$ mass NSs with $90\%$
credibility \cite{Miller:2021qha} and indicates that
$R_{1.4M_{\odot}}=11.34-13.23\,$km regarding $M=1.4M_{\odot}$ mass
NSs. Also a refinement of the NICER constraint was given in the
literature \cite{Ecker:2022dlg} which takes into account the heavy
black-widow binary pulsar PSR J0952-0607 which has mass $M=2.35\pm
0.17$ \cite{Romani:2022jhd}. We shall call this NICER II in the
$M-R$ plots.
\begin{figure}[h!]
\centering
\includegraphics[width=18pc]{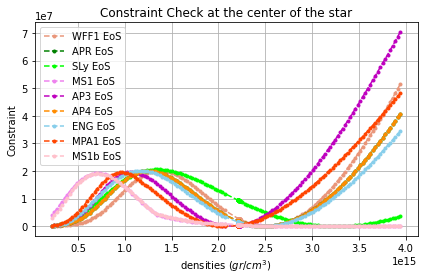}
\caption{The values of $\xi^2\phi^2$ as functions of the central
densities of the NS, for the center of the NS. As it can be seen
the constraint of Eq. (\ref{approximationaxion}) is well
satisfied.} \label{plot3}
\end{figure}
The NICER II indicates that the radius of a $M=1.4M_{\odot}$ NS
has to be $R_{1.4M_{\odot}}=12.33-13.25\,$km. For reading and
referral convenience, we present the NICER I and NICER II
constraints and constraints on the tidal deformability from the
GW170817 merger \cite{TheLIGOScientific:2017qsa} in Table
\ref{nicerconstraints}.
\begin{table}[h!]
  \begin{center}
    \caption{\emph{\textbf{NICER I AND NICER II Constraints for the radius of a $M=1.4M_{\odot}$ NS and Constraints for the Tidal Deformability from GW170817 }}}
    \label{nicerconstraints}
    \begin{tabular}{|r|r|}
 \hline
  NICER I & $11.34\,\mathrm{km}<R_{1.4M_{\odot}}<13.23\,\mathrm{km}$ \cite{Miller:2021qha} \\
  \hline
  NICER II & $12.33\,\mathrm{km}<R_{1.4M_{\odot}}<13.25\,\mathrm{km}$ \cite{Ecker:2022dlg} \\
  \hline
  GW170817 & $\tilde{\Lambda} \leq 800$ \cite{TheLIGOScientific:2017qsa} \\
  \hline
  GW170817 & $\Lambda(1.4M_{\odot}) \leq 800$ \cite{TheLIGOScientific:2017qsa} \\
  \hline
GW170817 & $\tilde{\Lambda}=245^{+453}_{-151}$ \cite{De:2018uhw} \\
  \hline
    \end{tabular}
  \end{center}
\end{table}
In Fig. \ref{plot1} we present the $M-R$ graphs for the axionic
NSs, regarding the Jordan frame masses and radii, considering the
following EoSs WFF1, SLy, APR, MS1, AP3, AP4, ENG, MPA1, MS1b and
we confront the $M-R$ graphs with the constraints NICER I
\cite{Miller:2021qha} and NICER II \cite{Ecker:2022dlg} (see Table
\ref{nicerconstraints}). From Fig. \ref{plot1} it is abundantly
clear that three EoSs pass the final test regarding axionic NSs,
the ENG (marginally pass), the AP3 (marginally pass too) and the
MPA1, which is the most optimal EoS. Now this is quite intriguing,
since in Ref. \cite{Odintsov:2023ypt}, which described  a
different context and perspective, the MPA1 EoS was also found to
be at the most optimal EoS phenomenologically. Specifically in
Ref. \cite{Odintsov:2023ypt} the scalar-tensor theories studied,
were basically inflationary attractors. In this work however we
consider an axionic scalar-tensor theory in which the axion is
considered to be in the oscillating era regime of its potential.
In simple words, the inflationary attractors describe viable
inflationary theories while in the present context, the axion
describes a dark matter theory, at least when the axion
oscillations commence and the potential is approximately a
quadratic one. Thus remarkably, two classes of conceptually
different scalar-tensor theories overlap for the MPA1 EoS. This is
quite intriguing and it is tempting to compare the inflationary
attractors with the axionic theory for the MPA1 EoS, which seems
to enjoy an elevated role among all the EoSs we considered. The
comparison is made in Fig. \ref{plot2}. As it can be seen, the
$M-R$ graph for the inflationary attractors and for the axionic
scalar-tensor theories are quite similar and in fact the axionic
scalar-tensor theory along with the Induced inflation and the
Higgs inflation cases, are well confronted with the NICER I and II
constraints, all the three enjoying an elevated position compared
to the other models.
\begin{table}[h!]
  \begin{center}
    \caption{\emph{\textbf{Axionic NSs vs CSIII for NS Masses $M\sim 1.6M_{\odot}$, $R_{1.6M_{\odot}}>10.68^{+0.15}_{-0.04}\,$km, for the SLy, APR, WFF1, MS1 and AP3 EoSs. The ''x'' denotes non-viability.}}}
    \label{tablecsiii16}
    \begin{tabular}{|r|r|r|r|r|r|}
     \hline
      \textbf{Model}   & \textbf{SLy EoS} & \textbf{APR EoS} & \textbf{WFF1EoS} & \textbf{MS1 EoS} & \textbf{AP3 EoS}
      \\  \hline
      \textbf{Axionic NSs Radii} & $R_{SLy}= 11.924\,$Km & $R_{APR}=  11.705\,$Km & $R_{WFF1}=
     10.892\,$Km
       & $R_{MS1}= x$ & $R_{AP3}= x$
\\  \hline
    \end{tabular}
  \end{center}
\end{table}
\begin{table}[h!]
  \begin{center}
    \caption{\emph{\textbf{Axionic NSs vs CSIII for NS Masses $M\sim
1.6M_{\odot}$, $R_{1.6M_{\odot}}>10.68^{+0.15}_{-0.04}\,$km, for
the AP4, ENG, MPA1 and MS1b. The ''x'' denotes non-viability.}}}
    \label{tablecsiii162}
    \begin{tabular}{|r|r|r|r|r|}
     \hline
      \textbf{Model}   & \textbf{AP4 EoS} & \textbf{ENG EoS} &
      \textbf{MPA1
      EoS} & \textbf{MS1b EoS}
      \\  \hline
      \textbf{Axionic NSs Radii} & $R_{AP4}= 11.705\,$Km & $R_{ENG}=  12.308\,$Km & $R_{MPA1}=
      12.871\,$Km
       & $R_{MS1b}= 14.902\,$Km
\\  \hline
    \end{tabular}
  \end{center}
\end{table}
Using the extracted data of our numerical analysis for the Jordan
frame masses and radii for the NSs, we also confronted these with
the CSI, CSII and CSIII constraints. We gathered the results in
several tables in the text for reading convenience. In Table
\ref{tablemaxmasses} we present the maximum masses of NSs that
 belong to the mass-gap region, for the corresponding EoSs that
 achieve this. Also in Tables \ref{tablecsi2}-\ref{tablecsi22} we
 confront the NSs radii with the CSI constraint, when NSs with
 masses $M\sim 2M_{\odot}$ are considered. Also in
Tables \ref{tablecsi14}-\ref{tablecsi142} we confront the NSs
radii with the CSI constraint, for NSs with masses $M\sim
1.4M_{\odot}$. Furthermore, in Tables
\ref{tablecsii14}-\ref{tablecsii142} we confront the NSs radii
with the CSII constraint, when NSs with masses $M\sim
1.4M_{\odot}$ are considered. Also in Tables
\ref{tablecsiii16}-\ref{tablecsiii162} we confront the radii of
NSs with masses $M\sim 1.6M_{\odot}$ with the constraint CSIII and
also in Tables \ref{tablecsiiimax}-\ref{tablecsiiimax2} we
confront again the radii of NSs with maximum masses, with the
constraint CSIII.
\begin{figure}[h!]
\centering
\includegraphics[width=18pc]{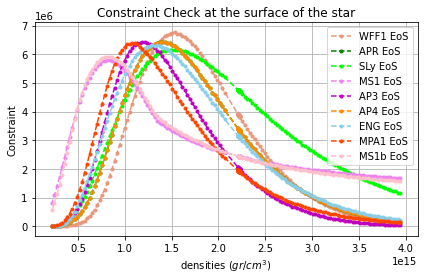}
\caption{The values of $\xi^2\phi^2$ as functions of the central
densities of the NS, for the surface of the NS. As it can be seen
the constraint of Eq. (\ref{approximationaxion}) is well
satisfied.} \label{plot4}
\end{figure}
Regarding the maximum masses, there exist several models that
predict a maximum mass inside the mass-gap region for the axionic
NSs but only the AP3 and the MPA1 and marginally the ENG EoS
predict a NS with mass in the mass-gap region. More importantly,
for these viable cases, the maximum mass is well below the causal
3 solar masses limit of GR. Recall that the causal mass GR limit
for static NSs is also respected for modified gravity theories too
\cite{Astashenok:2021peo}, and for GR it is
\cite{Rhoades:1974fn,Kalogera:1996ci},
\begin{equation}\label{causalupperbound}
M_{max}^{CL}=3M_{\odot}\sqrt{\frac{5\times
10^{14}g/cm^{3}}{\rho_u}}\, ,
\end{equation}
where $\rho_u$ stands for the reference density used to separate
the causal region and the low-density region, up to which the
overall EoS is known and the corresponding pressure is
$P_u(\rho_u)$. The causal EoS is,
\begin{equation}\label{causallimiteos}
P_{sn}(\rho)=P_{u}(\rho_u)+(\rho-\rho_u)c^2\, ,
\end{equation}
and for rotating NSs, the causal limit is,
\begin{equation}\label{causalrot}
M^{CL,rot}_{max}=3.89M_{\odot}\sqrt{\frac{5\times
10^{14}g/cm^{3}}{\rho_u}}\, ,
\end{equation}
but this does not concern us in this work, we quote it only for
completeness.
\begin{table}[h!]
  \begin{center}
    \caption{\emph{\textbf{Axionic NSs Maximum Masses and the Corresponding Radii vs CSIII, $R_{M_{max}}>9.6^{+0.14}_{-0.03}\,$km, for the SLy, APR, WFF1, MS1 and AP3 EoSs. The ''x'' denotes non-viability. }}}
    \label{tablecsiiimax}
    \begin{tabular}{|r|r|r|r|r|r|}
     \hline
      \textbf{Model}   & \textbf{APR EoS} & \textbf{SLy EoS} & \textbf{WFF1
      EoS} & \textbf{MS1 EoS} & \textbf{AP3 EoS}
      \\  \hline
      \textbf{Axionic NSs $M_{max}$} & $M_{APR}= 2.417\,M_{\odot}$ & $M_{SLy}= 2.272\, M_{\odot}$ & $M_{WFF1}= 2.320\,
M_{\odot}$ & $M_{MS1}= 3.175\,M_{\odot}$ & $M_{AP3}=
2.638\,M_{\odot}$
\\  \hline
\textbf{Axionic NSs Radii} & $R_{APR}= 10.579\,$Km & $R_{SLy}=
10.567\,$Km & $R_{WFF1}=
     9.911\,$Km
       & $R_{MS1}= 13.910\,$Km & $R_{AP3}= 11.359\,$Km
\\  \hline
    \end{tabular}
  \end{center}
\end{table}
\begin{table}[h!]
  \begin{center}
    \caption{\emph{\textbf{Axionic NSs Maximum Masses and the and the correspondent vs CSIII, $R_{M_{max}}>9.6^{+0.14}_{-0.03}\,$km, for the AP4, ENG, MPA1 and MS1b. The ''x'' denotes non-viability. }}}
    \label{tablecsiiimax2}
    \begin{tabular}{|r|r|r|r|r|}
     \hline
      \textbf{Model}   & \textbf{AP4 EoS} & \textbf{ENG EoS} &
      \textbf{MPA1
      EoS} & \textbf{MS1b EoS}
      \\  \hline
      \textbf{Axionic NSs $M_{max}$} & $M_{AP4}= 2.417\,M_{\odot}$ & $M_{ENG}= 2.492\, M_{\odot}$ & $M_{MPA1}= 2.771\,
M_{\odot}$ & $M_{MS1b}= 3.167\,M_{\odot}$
\\  \hline
\textbf{Axionic NSs Radii} & $R_{AP4}= 10.579\,$Km & $R_{ENG}=
11.011\,$Km & $R_{MPA1}=
      11.944\,$Km
       & $R_{MS1b}= 13.838\,$Km
\\  \hline
    \end{tabular}
  \end{center}
\end{table}
Now regarding the confrontation of the axionic NSs with the
constraints CSI and CSII, three EoSs are entirely excluded, the
WFF1, the MS1 and the MS1b EoSs. While in the case of the
constraint CSIII, the WFF1 is entirely incompatible with it.
Finally, we need to check the validity of the approximation
(\ref{approximationaxion}) for the center and the NS surface. This
is presented in Figs. \ref{plot3} and \ref{plot4} respectively. As
it can be seen, the constraint is well satisfied both at the
center and the surface of the NS.

\section*{Concluding Remarks}

In this work we considered the effects of an axionic scalar-tensor
theory on static NSs. Specifically, we considered the axionic
theory in the regime that the axion field oscillates around the
minimum of its scalar potential and cosmologically redshifts as
cold dark matter. Thus this scalar-tensor theory is somewhat
distinct from inflationary scalar-tensor theories. We presented
the essential features of the axion scalar-tensor theory and
demonstrated how this theory is basically a dark matter theory in
which the axion starts to behave as cold dark matter
post-inflationary. Specifically, when the Hubble rate of the
Universe becomes of the same order as the axion mass, the axion
begins coherent oscillations around the minimum of its potential
and redshifts as cold dark matter. The scalar-tensor theory we
chose to describe the axion basically takes into account the fact
that the axion oscillates around the potential minimum, so we took
into account this approximation in order to appropriately describe
the potential in this regime. We constructed the TOV equations for
this axionic theory, and we specified the initial conditions at
the center of the star. The we used a double shooting method in
order to find the optimal initial conditions at the center of the
star, for the scalar field value and the metric function, which
make the scalar field to smooth out to zero at the numerical
infinity. From our numerical analysis we calculated the Einstein
frame gravitational mass and radii for the NSs and accordingly we
found their Jordan frame counterparts using the formulas we
provided for the ADM mass and Jordan frame radius. The numerical
calculation was performed for nine distinct and physically
motivated EoSs, the WFF1, the SLy, the APR, the MS1, the AP3, the
AP4, the ENG, the MPA1 and the MS1b, using the piecewise
polytropic description for each EoS. From the resulting data we
constructed the $M-R$ graphs (Jordan frame quantities) and finally
we confronted the resulting NS phenomenology with the mainstream
of NS constraints available in the literature. Specifically we
used the NICER constraint and also a variant form of it
\cite{Ecker:2022dlg}, which we called NICER II, based on heavy
black-widow binary pulsar PSR J0952-0607 with mass $M=2.35\pm
0.17$ \cite{Romani:2022jhd}, which constraints the radius of an
$M=1.4M_{\odot}$ mass NS to have a radius
$R_{1.4M_{\odot}}=12.33-13.25\,$km. Also we used three extra
well-known constraints which we named CSI, CSII and CSIII
appearing in Table \ref{table0}. The resulting phenomenology was
deemed quite interesting for various reasons. Firstly, all the
viable EoSs for the axionic stars, predict a maximum mass in the
mass-gap region with $M_{max}>2.5M_{\odot}$, however with the
maximum mass being lower than the 3 solar masses causal EoS limit.
The three EoSs which are compatible with all the constraints we
imposed are the AP3, ENG and the MPA1, with the latter enjoying an
elevated role among all EoS. In fact the MPA1 EoS produces the
most well-fitted results which are compatible with all the
constraints. Also, the WFF1, MS1 and MS1b EoSs, are entirely
excluded from describing viable static NSs. Our results are
similar with the ones obtained when inflationary attractors NSs
are considered, and this intriguing fact made us compare the
axionic NSs with the inflationary attractors. As we demonstrated
the resulting $M-R$ graphs are quite similar, thus although the
two scalar-tensor theories have a different context, with the
axionic one corresponding to a dark matter theory (and its
approximations), the two theories produce quite similar NS
phenomenology. In conclusion, the MPA1 EoS is deemed
phenomenologically important, and it produces NSs with maximum
masses inside the mass gap region, below the 3 solar masses limit
though. Thus this EoS along with the scalar-tensor axionic model
and the inflationary attractor models will constitute a class of
models which can explain any future observation of NS with mass in
the mass-gap region but below 3 solar masses.

Finally, an important comment is in order, which must be explored
in a deeper way. Specifically, the NICER's results and constraints
are derived using ray tracing within the framework of general
relativity, but the external metrics of NSs in scalar-tensor
gravity differ from those in general relativity. Strictly
speaking, employing results from NICER or GW170817 can be
problematic. This issue, is very important from a phenomenological
point of view, and it was addressed in Refs.
\cite{Hu:2021tyw,Silva:2019leq,Silva:2018yxz}. We hope to discuss
this issue further in the future, in the context of inflationary
and axionic scalar-tensor theories of gravity.

\section*{Acknowledgments}

This research has been is funded by the Committee of Science of
the Ministry of Education and Science of the Republic of
Kazakhstan (Grant No. AP14869238).

\end{document}